\documentclass[conference]{IEEEtran}
\IEEEoverridecommandlockouts
\usepackage{cite}
\usepackage{amsmath,amssymb,amsfonts}
\usepackage{tablefootnote}
\usepackage[format=plain, labelfont={bf,it}]{caption} 
\usepackage{algorithmic}
\usepackage{graphicx}
\usepackage{textcomp}
\usepackage{xcolor}
\usepackage{makecell}
\usepackage{listings}
\usepackage{hhline}

\usepackage{url}
\def\BibTeX{{\rm B\kern-.05em{\sc i\kern-.025em b}\kern-.08em
    T\kern-.1667em\lower.7ex\hbox{E}\kern-.125emX}}
    
\DeclareCaptionFormat{listing}{\rule{\dimexpr\textwidth+17pt\relax}{0.4pt}\vskip1pt#1#2#3}
\begin{document}
\bstctlcite{IEEEexample:BSTcontrol}

\title{Porting incompressible flow matrix assembly to FPGAs for accelerating HPC engineering simulations}

\author{\IEEEauthorblockN{Nick Brown}
\IEEEauthorblockA{\textit{EPCC, University of Edinburgh} \\
Edinburgh, UK \\
n.brown@epcc.ed.ac.uk}
}

\maketitle

\begin{abstract}
Engineering is an important domain for supercomputing, with the Alya model being a popular code for undertaking such simulations. With ever increasing demand from users to model larger, more complex systems at reduced time to solution it is important to explore the role that novel hardware technologies, such as FPGAs, can play in accelerating these workloads on future exascale systems. 

In this paper we explore the porting of Alya's incompressible flow matrix assembly kernel, which accounts for a large proportion of the model runtime, onto FPGAs. After describing in detail successful strategies for optimisation at the kernel level, we then explore sharing the workload between the FPGA and host CPU, mapping most appropriate parts of the kernel between these technologies, enabling us to more effectively exploit the FPGA. We then compare the performance of our approach on a Xilinx Alveo U280 against a 24-core Xeon Platinum CPU and Nvidia V100 GPU, with the FPGA significantly out-performing the CPU and performing comparably against the GPU, whilst drawing substantially less power. The result of this work is both an experience report describing appropriate dataflow optimisations which we believe can be applied more widely as a case-study across HPC codes, and a performance comparison for this specific workload that demonstrates the potential for FPGAs in accelerating HPC engineering simulations.
\end{abstract}


\section{Introduction}
Designing safer cars, more energy efficient aircraft, and improved turbines are examples of the benefits that engineers and society gains from computer simulation. Due to the computational requirements involved in such simulations, these workloads have exploited HPC for many years. However, driven by ever increasing user ambition, in part accelerated by the push to de-carbonise and produce more efficient engineering solutions, engineers require the ability to model much larger, more complex systems on future supercomputers.

Alya \cite{vazquez2016alya} is a multi-physics simulation code developed at Barcelona Supercomputing Centre (BSC) which, since its inception in 2004, has proven popular for use in studying a wide variety of engineering systems. Alya follows a plugin architecture, where a series of modules are provided, each representing a single set of Partial Differential Equations (PDE) for a given physical model. The developers of Alya are interested in exploring the potential role of novel hardware architectures to help understand the most appropriate future HPC technologies required to meet the next generation of engineering challenges. One such technology, popular in other fields but yet to gain widespread acceptance in HPC, is that of Field Programmable Gate Arrays (FPGAs) which provide a large number of re-configurable logic blocks sitting within a sea of configurable interconnect. Recent years have seen significant advances made by FPGA vendors which aim to address the limited hardware capabilities and esoteric programming environments that historically limited uptake of this technology for HPC workloads \cite{brown2020weighing}. For instance FPGAs now also contain, as standard, very fast on-chip memory, DSP slices for accelerating floating point arithmetic, and high bandwidth external connections. Consequently FPGAs are more accessible for HPC workloads and developers than ever before, and a key question is whether they can compliment existing CPU and GPU hardware in future exascale machines.

There have been numerous efforts porting incompressible flow calculations to FPGAs. The Himeno benchmark has been used to demonstrate the suitability of FPGAs for this workload \cite{fujita2019parallel}, and other studies exploring methods for solving equations arising from these systems \cite{oyarzun2021fpga} \cite{karp2021high}. By contrast in this work we focus top down on one specific real-world HPC simulation application, namely Alya, and are driven from the perspective of the application developer by the bottlenecks in that code. As described in Section \ref{sec:alya}, we focus on building the matrix for the incompressible flow solver as it is this kernel that is responsible for significant amount of application level runtime and our hypothesis that the dataflow architecture can help ameliorate the memory bound nature of the code. The specifics of the calculations and code structure are unique to Alya and hence this is an interesting case-study around optimising such structures on FPGAs and benefiting Alya. 

In this paper we explore the acceleration of Alya's incompressible flow matrix assembly kernel \cite{cajas2018fluid}, which is a foundational component of many run configurations and accounts for a significant fraction of the overall model runtime. Aimed in part as an experience report, we explore the actions undertaken to optimise the kernel for FPGAs and achieve good performance for this specific workload, with a view that these can then apply more widely when accelerating HPC kernels on FPGAs. The rest of this paper is structured as follows, in Section \ref{sec:bg} we explore the background to this work, describing the current state of the art FPGA tool chains that are used in this work, and related work in the field of HPC. This is then followed by a detailed description of the Alya model, and the specific kernel of interest, and lastly a description of the hardware configurations used in this work is presented. Section \ref{sec:kernel} focuses on the development and optimisation of our FPGA kernel for matrix assembly, exploring the steps required to achieve good performance before describing sharing parts of the workload between the FPGA and CPU to make best use of each's capabilities. Section \ref{sec:performance} then explores performance and power characteristics of our FPGA approach against a 24-core Xeon Platinum CPU and Nvidia V100 GPU, before we draw conclusions and discuss further work in Section \ref{sec:conclusions}

\section{Background}
\label{sec:bg}
High Level Synthesis (HLS) has empowered programmers to write C or C++ code for FPGAs and for this to be translated into the underlying Hardware Description Language (HDL). As a key component of Xilinx's Vitis toolchain \cite{vitis}, HLS avoids the need to write code at the HDL level for FPGAs, significantly improving productivity, opening up the programming of FPGAs to a much wider community, and enabling us to focus more at the algorithmic dataflow level. Furthermore, Vitis also automates the process of integration with the wider on-chip infrastructure such as memory controllers and interconnects, as well as making emulation and performance profiling more convenient. In this approach the programmer decorates their C++ device code with Xilinx's bespoke pragma style hints to drive the tooling, and host code is written in OpenCL.

However, HLS is not a silver bullet and whilst it has made the physical act of programming FPGAs much easier, one must still target the most appropriate kernels for acceleration on FPGAs \cite{brown2020exploring} and recast their Von-Neumann style algorithms into a dataflow style \cite{koch2016fpga} to obtain best performance. Whilst there have been numerous activities exploring the role of FPGAs for HPC workloads \cite{yang2019fully}, \cite{brown2020exploring}, \cite{lfricfpga}, and a number of successes, many struggle in-terms of performance against the latest CPUs and GPUs especially if the HLS code has not been fully optimised for the dataflow architecture. Whilst tuning application codes to suit target hardware is not new for HPC programmers, for instance it is well accepted that moving from CPU to GPU architectures requires code level changes, the dataflow rather than Von-Neumann nature of FPGAs means that typically far more substantial changes are required which can involve a fundamental recast of the algorithm. Combined with the long build time for FPGAs this can be a time-consuming process, and-so application level studies for high performance codes such as the one contained in this paper are worthwhile for exploring and disseminating appropriate techniques and code structures.

To most effectively exploit FPGAs it is also important to select a kernel which will benefit from execution on a dataflow architecture. Due to the high level of raw floating point performance provided by GPUs and to a lesser extent the latest generation of CPUs, if a kernel is compute bound then realistically the FPGA will likely struggle to compete. However if kernel is bound by other factors, such as being memory bound or bound by micro-architecture issues, then potentially moving to the dataflow model of the FPGA will be beneficial. This is because, without a black-box microarchitecture imposed by other technologies, the programmer has more opportunity to tailor their kernel and data accesses to completely suit the application in question and ameliorate issues found on other more general purpose architectures.

\subsection{Alya}
\label{sec:alya}
Alya \cite{vazquez2016alya} is a high performance computational mechanics code used to solve complex coupled multi-physics, multi-scale, and multi-domain problems. Written in Fortran and aimed at engineering, Alya handles many physical systems including incompressible and compressible flows, non-linear solid mechanics, chemistry, particle transport, multiphase problems, heat transfer, turbulence modelling, and electrical propagation \cite{vazquez2014alya}. Furthermore, Alya is one of two CFD codes of the Unified European Applications Benchmark Suite (UEBAS) \cite{bull7unified} as well as the PRACE Accelerator benchmark suite \cite{hautreuxdescription}.

A run configuration is made up of modules, each of which implements specific functionality representing a single set of Partial Differential Equations (PDE) for a given physical model. In this work we are focusing on the \emph{nastin} module which models incompressible flow and is a key component of many run configurations. In this module the building of the matrix, which is used to solve Navier-Stokes equations, is a costly operation, representing 64\% of the overall model runtime for the \emph{Sphere 16M} benchmark (see Table \ref{tbl:benchmark_details}). When profiling with Intel VTune we found that the kernel was stalling 32\% of the time due to memory accesses and 11\% due to other microarchitecture core-bound issues. Therefore an important question is whether on the FPGA, by designing memory access in a bespoke manner, we can ameliorate such issues imposed by the general purpose CPU architecture.

Listing \ref{lst:matrix_assembly} sketches the matrix assembly code in the subroutine \emph{matrix\_assembly}, where \emph{veloc} and \emph{coord} are real number inputs, with \emph{lnods} providing a mapping between the data required each element and the indexes in these arrays where it is located. Each subroutine called by the code of Listing \ref{lst:matrix_assembly} builds intermediate values which are made available to subsequent calculations for the current element. These temporaries are mainly a mix of 1D and 2D arrays, although there are some exceptions where they are scalar real numbers. 

The structure of the \emph{calculate\_cartesian\_derivatives} subroutine of Listing \ref{lst:matrix_assembly} is typical of the computational routines, where typically there is an overarching loop over \emph{PGAUS\_SIZE}, which is the number of Gauss points, and then nested loops on a mixture of \emph{PNODE\_SIZE} (the number of nodes) and/or \emph{NUM\_DIMS} (the number of dimensions). All calculations are double precision floating point.


\begin{lstlisting}[frame=lines,caption={Sketch of matrix assembly code and illustrates the loop structure of computational routines},label={lst:matrix_assembly}, numbers=left]
subroutine matrix_assembly(...)
 do e=1, number_elements
  call calculate_transients(veloc, coord, lnods, elvel, elcod)
  call calculate_cartesian_derivatives(..., elcod, gpcar, gpvol)
  call calculate_gauss_point_values(..., elvel, gpcar, gpvel, gpadv, gprhs, gpgve)
  call calculate_tau_and_tim(..., gpvol, ..., eldtrho, elmurho)
  call calculate_element_matricies(..., gpcar, gpadv, agrau, wgrgr, ...)
  call calculate_convective_term_and_RHS(agrau, gpcar, gpvol, gpvel, gprhs, gpgve, ..., elauu, elrbu)
  call calculate_viscous_term(wgrgr, gpvol, elvel, gpcar, elauu, elrbu, ..., elrbu)
  call perform_assembly_in_global_system(elrbu, eldtrho, elmurho, lnods, rhsid, dt_rho_nsi, mass_rho_nsi)
 end do
end subroutine matrix_assembly

subroutine calculate_cartesian_derivatives(...)
 do igaus=1, PGAUS_SIZE
  do k=1, PNODE_SIZE
   ...
   do i=1, NUM_DIMS
    ...
   end do
  end do
  ...
  do j=1, PNODE_SIZE
   ...
  end do
 end do
end subroutine calculate_cartesian_derivatives
\end{lstlisting}

Data input is handled by the \emph{calculate\_transients} subroutine of Listing \ref{lst:input_output}, which illustrates  the loading of data from the \emph{veloc} and \emph{coord} arrays into internal \emph{elvel} and \emph{elcod} 2D arrays for a specific element, unpacking each node of an element (there are 4 nodes per elements in the benchmarks used in this work), and then each dimension of a node (there are three) with the mappings directed by the \emph{lnods} index array. 

The variables \emph{rhsid}, \emph{dt\_rho\_nsi}, and \emph{mass\_rho\_nsi} are outputs of the matrix assembly kernel, where the \emph{perform\_assembly\_in\_global\_system} subroutine of Listing \ref{lst:input_output} unpacks internal temporary calculated values and accumulates these into output arrays. Similarly to the input values, the mapping between the array location for results is determined by the \emph{lnods} mapping array.

\begin{lstlisting}[frame=lines,caption={Sketch of input and output subroutines for matrix assembly calculation},label={lst:input_output}, numbers=left]
subroutine calculate_transients(veloc, coord, lnods, elvel, elcod)
 real(rp), intent(in) :: veloc(:,:), coord(:,:)
 integer, intent(in) :: lnods(:,:)
 real(rp), intent(out) :: elvel(:,:), elcod(:,:)

 integer :: ipoin
 do i=1, PNODE_SIZE
  ipoin=lnods(e, i)
  do j=1, NUM_DIMS
   elvel(j,i) = veloc(j,ipoin)
   elcod(j,i) = coord(j,ipoin)
  end do
 end do
end subroutine calculate_transients

subroutine perform_assembly_in_global_system(elrbu, eldtrho, elmurho, lnods, rhsid, dt_rho_nsi, mass_rho_nsi)
 real(rp), intent(in) :: elrbu(:,:), eldtrho(:), elmurho(:)
 integer, intent(in) :: lnods(:,:)
 real(rp), intent(inout) :: rhsid(:,:), dt_rho_nsi(:), mass_rho_nsi(:)
  
 integer :: ipoin
 do i = 1, PNODE_SIZE
  ipoin=lnods(e, i)
  do j=1, NUM_DIMS
   rhsid(idime, ipoin) = rhsid(idime, ipoin) + elrbu(j,i)
  end do
  dt_rho_nsi(ipoin) = dt_rho_nsi(ipoin) + eldtrho(j)
  mass_rho_nsi(ipoin) = mass_rho_nsi(ipoin) + elmurho(j)
 end do
end subroutine perform_assembly_in_global_system
\end{lstlisting}

Table \ref{tbl:matrix_assembly_details} details the percentage time of each matrix assembly subroutine in the \emph{Sphere 100K} benchmark (see Table \ref{tbl:benchmark_details}), along with the number of double precision floating point operations required. In total there are 7052 double precision floating point operations required per element, with the most substantial routines being that of \emph{calculate\_convective\_term\_and\_RHS} and \emph{calculate\_viscous\_term} which also account for the largest percentages of overall contribution time.

\begin{table}[h]

 \centering
\begin{tabular}{ | c c c | }
\hline
\textbf{Routine} & \textbf{Time} & \textbf{FLOPs per element} \\ \hline
calculate\_transients & 3.2\% & 0 \\
calculate\_cartesian\_derivatives & 5.4\% & 664 \\
calculate\_gauss\_point\_values & 8.9\% & 400 \\
calculate\_tau\_and\_tim & 2.1\% & 76 \\
calculate\_element\_matricies & 11\% & 416 \\
calculate\_convective\_term\_and\_RHS & 40\% & 3936 \\
calculate\_viscous\_term & 26\% & 1540 \\
perform\_assembly\_in\_global\_system & 3.4\% & 20 \\
\hline
\end{tabular}
\caption{Performance details of each component making up matrix assembly kernel}
\label{tbl:matrix_assembly_details}
\end{table}

Table \ref{tbl:benchmark_details} provides an overview of the benchmarks used throughout this work, detailing their element sizes, number of nodal points (this determines the size of the input and output arrays, with corresponding nodal point entries for each element determined by the \emph{lnods} mapping), and associated input and output data sizes (which for the FPGA must be transferred over PCIe). The two benchmarks which are stared, \emph{Sphere 100K} and \emph{Sphere 32M} are, unlike the other four benchmarks, not part of the Alya standard benchmark suite but instead developed from the Sphere benchmark. This is because there is a large element size difference in the suite between small and large benchmarks. Therefore, these additional two benchmark configurations help provide a more complete picture in our performance evaluations across a wider range of element sizes.

\begin{table}[h]

 \centering
\begin{tabular}{ | c c c c c | }
\hline
\textbf{Benchmark} & \makecell{\textbf{Number} \\ \textbf{of elements}}& \makecell{\textbf{Number of} \\ \textbf{nodal points}} & \makecell{\textbf{Input} \\ \textbf{data size}}& \makecell{\textbf{Output} \\ \textbf{data size}}\\ \hline
Cylinder 2D & 1200 & 1280 & 0.16MB & 0.06MB \\
Venturi 2D & 4200 & 4371 & 0.56MB & 0.20MB \\
Elbow & 26410 & 5682 & 1.05MB & 0.26MB \\
Sphere 100K\tablefootnote[1]{This size is not an official benchmark and was created based on the Sphere benchmark to increase the range of element sizes being evaluated} & 100000 & 15768 & 3.33MB & 0.72MB \\
Sphere 16M & 16677400 & 2876880 & 584MB & 132MB \\
Sphere 32M\footnote[1]{text} & 32677400 & 5753760 & 1157MB & 263MB \\
\hline
\end{tabular}
\caption{Details of Alya benchmark suite configurations selected as a basis of evaluating this work (see footnote for stared configurations)}
\label{tbl:benchmark_details}
\end{table}

\subsection{Hardware setup}

For the runs contained in this paper we use a Xilinx Alveo U280 which contains an FPGA chip with 1.08 million LUTs, 4.5MB of on-chip BRAM, 30MB of on-chip URAM, and 9024 DSP slices. This PCIe card also contains 8GB of High Bandwidth Memory (HBM2) and 32GB of DDR DRAM on the board, although for this work we use the HBM2 exclusively. The FPGA card is hosted in a system with a 26-core Xeon Platinum (Skylake) 8170 CPU. Codes for the Alveo are built with Xilinx Vitis framework version 2021.1.

We also compare against the code running on a 24-core Xeon Platinum (Cascade Lake) 8260M CPU, and Nvidia Tesla V100 GPU. On the CPU the code has been parallelised via OpenMP (using GCC 8.3) and on the GPU it uses OpenACC (using Nvidia compiler version 20.9). All reported numbers are averaged over three runs. More details are in Appendix \ref{sec:ad}.

\section{FPGA kernel development and optimisation}
\label{sec:kernel}
\begin{figure}[h]
\centering
\includegraphics[scale=0.45]{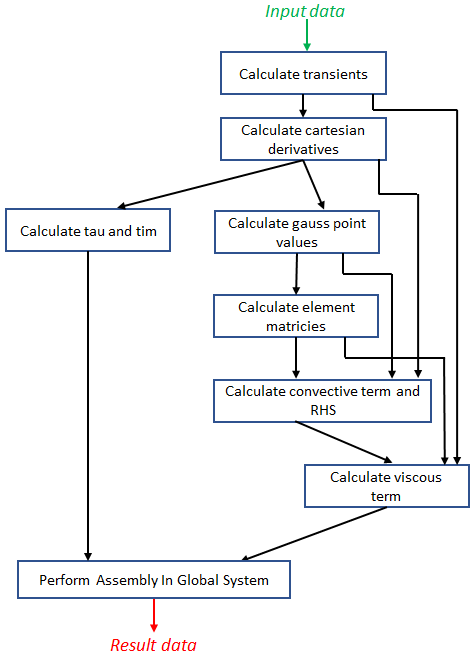}
\caption{Initial dataflow design of Alya incompressible flow matrix assembly engine on the FPGA.}
\label{fig:initial_df}
\end{figure}

From the sketch of the matrix assembly code in Listing \ref{lst:matrix_assembly} it can be seen that the constituent calculations generate intermediate data for a specific element, which is then used as an input to subsequent calculations for this same element. We believed that such an approach would suit the approach of a dataflow design, where each part is running concurrently and streaming these intermediate values from the producer to the consumer. Rewriting the kernel in C++ for the device, and writing OpenCL on the host to interface with the existing Alya code-base, our initial dataflow design is illustrated in Figure \ref{fig:initial_df}, where the blocks represent functions running concurrently for an element with data streaming between them. Numerous streams have multiple consumers, for instance \emph{elvel} is produced by \emph{calculate\_cartesian\_derivatives}, and consumed by both \emph{calculate\_gauss\_point\_values} and \emph{calculate\_viscous\_term}. In these instances, whilst it is not shown in Figure \ref{fig:initial_df} due to brevity, we have an additional data replication dataflow stage which accepts the stream to be replicated as an input and produces an array of streams as an output, each of which is consumed by a different dataflow stage and each cycle the replication utility functionality will read from the source stream and write to each target stream. 


\subsection{Optimizing the computational engine}

Table \ref{fig:kernel_performance} illustrates performance, for the Sphere 100K benchmark, for different versions of our FPGA kernel as we optimised it. We include the kernel only execution time, time taken for data transfers between the host and FPGA, and total execution time which is a combination of these first two measures. For reference the table also includes performance of the code running on a single core of the CPU and the Xeon's entire 24 cores. For each FPGA configuration in Table \ref{fig:kernel_performance} we provide the percentage of DSP and LUT resources required on a Super Logic Region (SLR) basis, with three SLRs present on the U280. All FPGA versions are running at a clock frequency of 300MHz, the U280 default. Details of our initial design are reported in Table \ref{fig:kernel_performance} as \emph{Initial FPGA dataflow design}, and achieving only 0.39\% the performance of the 24-core CPU it can be seen that significant optimisations were required! We have focused on using the Sphere 100K benchmark, a medium problem size, as a vehicle for driving optimisations due to the excessive initial execution time on the FPGA, with the wider range of benchmarks considered in Section \ref{sec:performance}. Optimising for this medium sized benchmark also directly translates into the other problem sizes.

\begin{table*}[h]

 \centering
\begin{tabular}{ | c c c c c c c | }
\hline
\textbf{Description} & \makecell{\textbf{Total execution} \\ \textbf{time (ms)}} & \makecell{\textbf{Kernel execution} \\ \textbf{time (ms)}} & \makecell{\textbf{Host-FPGA data} \\ \textbf{transfer time (ms)}} & \makecell{\textbf{\% CPU} \\ \textbf{performance}} & \makecell{\textbf{\% SLR DSP} \\ \textbf{usage}} & \makecell{\textbf{\%SLR LUT} \\ \textbf{usage}} \\\hline
1 core of Xeon CPU & 351.05 & 351.05 & - & - & - & -\\
24 cores Xeon CPU & 61.72 & 61.72 & - & - & - & - \\
Initial FPGA dataflow design & 15714.99 & 15701.78 & 13.21 & 0.39\% & 15\% & 13\%\\
Optimised II of loops & 1508.60 & 1495.62 & 12.98 & 4.09\% & 91\% & 40\%\\
Brought elements loop into DF functions & 293.21 & 279.79 & 13.42 & 21.05\% & 91\% & 44\%\\
Refactored code into engine & 284.04 & 270.71 & 13.33 & 21.73\% & 97\% & 48\%\\
\hline
\end{tabular}
\caption{Performance of FPGA matrix assembly engine with Sphere 100K benchmark running on a Xilinx Alveo U280 as kernel level optimisations were applied. Also included is performance of code running on a Xeon Platinum (Cascade Lake) 8260M CPU for comparison.}
\label{fig:kernel_performance}
\end{table*}

One of the major reasons for this initial poor performance was that our dataflow design was susceptible to deadlock, where if streams were written to in a different order than they were consumed from then the FPGA could hang. Furthermore, HLS automatically reorders stream writes and reads based upon dependencies in the code, so carefully laying out and manually ordering the stream accesses in code does not necessarily solve the problem. Therefore, whilst we had pipelined as many loops as we could, there were a number where it was not possible to pipeline due to this deadlocking and instead sub-loops were pipelined. This meant that the functions in our dataflow machine were rather asymmetrical in their pipelined behaviour, where some were pipelined on the outer \emph{PGAUS} loop, whereas others only pipelined on inner nested loops. As such it meant that, from a performance perspective, some stages had a tendency to stall preceding or subsequent stages, effectively reducing the amount of concurrency present in the design.


Whilst one approach would be to leverage HLS's \emph{protocol} pragma, which enforces HLS not reordering contained code, this would still require careful manual ordering of accesses in code by the programmer. Instead, we increasing the stream FIFO depth in the \emph{HLS STREAM} pragma, and also disallowed streams jumping ahead of subsequent stages. This involved routing streams through subsequent dataflow functions, irrespective of whether they utilised that streaming data or not in their calculations, and this approach follows Xilinx best practice \cite{vitis_bestpractice}. 

However it was not just the issue of deadlocking that was limiting the pipelining of loops. In HLS when a loop is pipelined then all contained inner loops must be completely unrolled. In our code there are numerous nested loops and unrolling all these could result in significant DSP usage on the FPGA. Under certain conditions, for instance if there are no statements between outer and inner loops, then HLS can merge these automatically when the pipeline pragma is applied to the inner loop. In this manner, for some functions in our matrix assembly engine, we were able to limit the amount of unrolling required whilst still achieving a pipelined outer loop. However for others this was not possible and the increased DSP usage had to be accepted. 

Furthermore, the \emph{calculate\_convective\_term\_and\_RHS} and \emph{calculate\_viscous\_term} functions contained a spatial dependency which was critical to fix because, as illustrated in Table \ref{tbl:matrix_assembly_details}, these account for a large portion of the overall matrix assembly runtime. A spatial dependency is where the calculations involved at one cycle depend on previous calculations which might not yet have completed. In our specific case there were accumulations on data, and these double precision additions required seven cycles to complete. Therefore the pipelined loop is limited to an Initiation Interval (II), the number of cycles before the next value can start to be processed, of seven because this number of cycles must elapse before the next value can start to be processed due to the dependence on the previously accumulated value.

Listing \ref{lst:spatial_dependency} illustrates a sketch of the algorithm that was causing these issues, where these two functions are different from the others because they only generate a single, accumulated, value per element, rather than a value for each loop iteration over \emph{PGAUS}. This property provided us with some flexibility when it came to refactoring the code to remove the spatial dependency. The code builds the \emph{elauu} internal 2D array (each dimension of size \emph{PNODE\_SIZE * NUM\_DIMS}), which is streamed out to the next dataflow stage at the end. There are in-fact two spatial dependencies, firstly between outer loop iterations of the \emph{PGAUS} loop, where the same elements of the \emph{elauu} variable are accumulated from one outer iteration to the next, and secondly between line 10 and the loop of lines 12 to 15.

\begin{lstlisting}[frame=lines,caption={Sketch of algorithmic structure which resulted in a spatial dependency},label={lst:spatial_dependency}, numbers=left]
for (int igaus=0;igaus<PGAUS_SIZE;igaus++) {
 ...
 for (int inode=0;inode<PNODE_SIZE;inode++) {
  ...
  for (int idime=0;idime<NUM_DIMS;idime++) {
   int idofv = inode * NUM_DIMS + idime;
    
   for (int jnode=0;jnode<PNODE_SIZE;jnode++) {
    int jdofv = jnode * NUM_DIMS + idime;
    elauu[jdofv][idofv] = elauu[jdofv][idofv] + ...;

    for (int jdime=0;jdime<NUM_DIMS;jdime++) {
     int jdofv = jnode*NUM_DIM+jdime;
     elauu[jdofv][idofv] = elauu[jdofv][idofv] ...;
    }
   }
   ...
  }
  ...
 }
}
elauu_stream.write(elauu);
\end{lstlisting}

It is possible to address the first spatial dependency via the \emph{dependence} HLS pragma, because whilst the HLS compiler can not guarantee there are enough cycles between one outer loop iteration and the next due to the dynamic indexes being calculated, this is obvious to the programmer. The second spatial dependency between line 10 and the loop of lines 12 to 15 is more difficult to remove. In its current form the algorithm will not provide the required number of cycles between subsequent accumulations and-so had to be refactored to remove this dependency and enable pipelining with initiation interval of one. 

Listing \ref{lst:fixed_spatial_dependency} sketches the refactored algorithm, where the base calculation is performed into variable \emph{c0} at line 13 and the \emph{special\_jdofv} selects those subset of elements which require additional calculations beyond those in \emph{c0}. In this manner the accesses for each element of the array \emph{elauu} at line 15 or 17, depending if it is a \emph{special\_jdofv} index or not, represent the entire calculation required for this value for the \emph{PGAUS} outer loop, rather than requiring accumulation in the algorithm. Instead of accumulating over \emph{PGAUS} and generating one single value per element, the algorithm streams out each value generated for the \emph{PGAUS} loop, and a subsequent dataflow stage is added to accumulate these. 

These optimisations enabled data to flow more effectively between the dataflow stages, improving the overall concurrency of our design and the performance benefits are illustrated by the entry \emph{Optimised II of loops} in Table \ref{fig:kernel_performance}. It can be seen that this significantly improves the performance of our kernel, with it running over 10 times faster, but at a higher resource usage cost, where DSP usage has increased from 15\% of an SLR in the previous version to 91\% now, and LUT usage has also increased, albeit at a lower rate. However the kernel was still only achieving around 4\% of the CPU's performance, so clearly there were still further optimisation opportunities.

\begin{lstlisting}[frame=lines,caption={Sketch of refactored algorithm to address spatial dependency and deliver an initiation interval of 1 },label={lst:fixed_spatial_dependency}, numbers=left]
for (int igaus=0;igaus<PGAUS_SIZE;igaus++) {
#pragma HLS PIPELINE II=1
 ...
 for (int inode=0;inode<PNODE_SIZE;inode++) {
  ...
  for (int idime = 0; idime<NUM_DIMS;idime++) {
   int idofv = inode * NUM_DIMS + idime;
   for (int jnode = 0; jnode<PNODE_SIZE;jnode++) {
    int special_jdofv = jnode * NUM_DIMS + idime;

    for (int jdime = 0; jdime<NUM_DIMS;jdime++) {
     int jdofv = jnode*NUM_DIMS+jdime;
     REAL_TYPE c0=...
     if (jdofv == special_jdofv) {
      elauu[jdofv][idofv] = c0 + ... 
     } else {
      elauu[jdofv][idofv] = c0;
     }
    }
   }
  }
 }
 elauu_stream.write(elauu);
}
\end{lstlisting}

At this point the loop over the number of elements was outside the dataflow pragma, meaning that between each element the dataflow stages had to shut down and restart which resulted in overhead. To address this we brought the elements loop inside each of the dataflow stages, ensuring that each dataflow stage could run continually from one element to the next. Using the same technique as described above, whilst only the \emph{PGAUS} loops were pipelined, as we did not have enough DSP slices to fully unroll each \emph{PGAUS} loop, we ensured that the loop over the number of elements was fused by HLS with this {PGAUS} loop, effectively meaning that both were pipelined. Moving the elements loop inside each dataflow region and ensuring that this loop fusion occurred sped the kernel up over five times, as reported by \emph{Brought elements loop into DF functions} in Table \ref{fig:kernel_performance}, although the FPGA was still only achieving around 21\% of the CPU's performance.

\begin{figure}[h]
\centering
\includegraphics[scale=0.45]{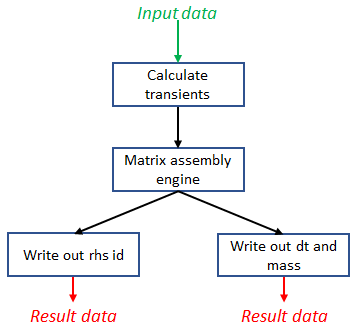}
\caption{Illustration of splitting out the Alya incompressible flow matrix assembly engine from the external data processing, such that the loading of input data and accumulation and storing of results are external with HLS streams between these components.}
\label{fig:split_out_df}
\end{figure}

In preparation for the optimisations that are discussed next in Section \ref{sec:improve_read_write}, as well as improving the code generally, we decided that it would be best to split out the reading and writing of input and output data, the \emph{calculate\_transients} and \emph{perform\_assembly\_in\_global\_system} functions from the rest of the matrix assembly computational engine. Effectively this meant that the engine would continually accept streams of element input data, process these, and then stream out resulting data. This produced two dataflow regions, one for the matrix assembly engine and the other for the reading and writing of input and output data. 

\begin{figure}[h]
\centering
\includegraphics[scale=0.45]{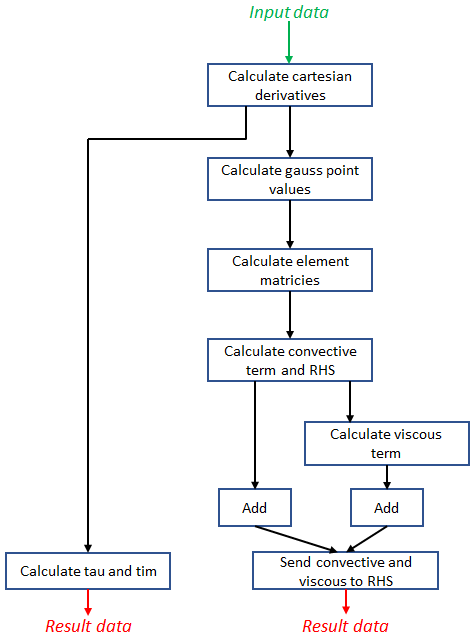}
\caption{Revised matrix assembly engine dataflow design based on optimisations discussed in this section, most notably streams are always routed through subsequent stages regardless of the consumption of data from them.}
\label{fig:revised_df}
\end{figure}

The top level dataflow region is illustrated in Figure \ref{fig:split_out_df}, with Figure \ref{fig:revised_df} illustrating the matrix assembly computational engine dataflow machine design which is nested. The engine of Figure \ref{fig:revised_df} incorporates the improvements previously discussed in this section where, for instance, it can be seen that streams no long skip stages, instead being routed through subsequent steps to avoid deadlock, and also additional \emph{add} stages towards the end as part of the refactoring of the convective and viscous calculations. At the same time, again following best practice \cite{vitis_bestpractice}, we modified external HBM2 memory accesses so that these were 512-bits wide by packing and unpacking data. This refactoring slightly improved performance, as represented by the \emph{Refactored code into engine} entry of Table \ref{fig:kernel_performance}, with the main objective being to improve the structure of the code for the next set of optimisations.

\subsection{Reducing external data access overhead}
\label{sec:improve_read_write}
At this point, whilst we had spent considerable time optimising the computational engine of our design to ensure that it could continually stream data, performance was still falling significantly short of that delivered by the 24-core Xeon Platinum CPU. An important question was whether the design was being most efficiently fed with input data from, and results being delivered to, the HBM2 external memory. Put simply, whether the engine was stalling due to a lack of input data or stalling because it was unable to stream out results due to overhead on writing. From Listing \ref{lst:input_output} it can be seen that the external memory access pattern is irregular, where there is an index held in the \emph{lnods} array which determines the locations to read from for the \emph{veloc} and \emph{coord} arrays, and write to for the \emph{rhsid}, \emph{dt\_rho\_nsi}, and \emph{mass\_rho\_nsi} arrays. This resulted in two major disadvantages for performance, firstly there was a spatial dependency for result data, as the accumulation of array elements could be followed by an accumulation into that same element for the next node, resulting in an initiation interval of 7 imposed by the tooling, and secondly accesses to external memory were not contiguous.

For each separate external memory access the HLS tooling has to add an explicit read request for input data and write response for output data, both costing 69 cycles. For contiguous external memory accesses the compiler can fuse accesses together, effectively meaning that there is one of these expensive operations for many individual accesses. However, due to the irregular access pattern this was not possible, making the pipelines very deep. This is one of the reasons why bringing the loop over the number of elements in Section \ref{sec:kernel} was so beneficial, as it enabled these pipelines to be filled for longer. However, such non-contiguous memory access patterns can significantly reduce external memory access performance \cite{brown2019s} regardless, and-so this was important to address.

\begin{figure}[h]
\centering
\includegraphics[scale=0.35]{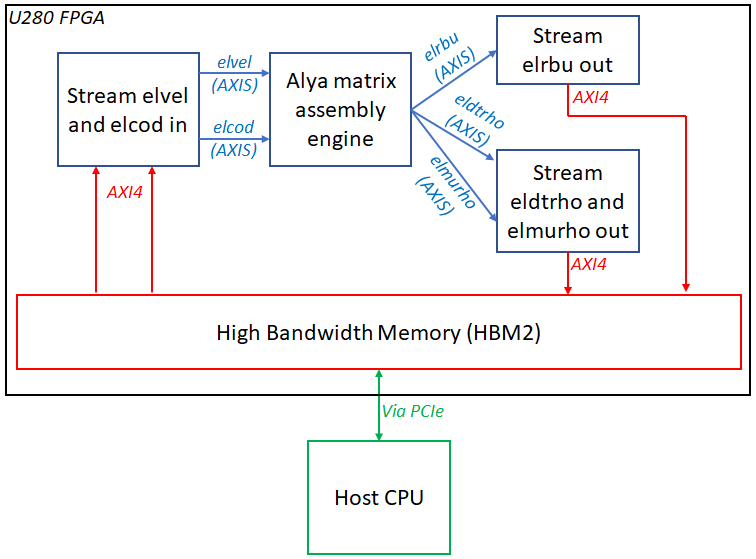}
\caption{Architectural view of how the host, HBM2, and IP blocks interact with the streaming design, where chunks of data in the format required for the Alya incompressible flow matrix assembly engine are streamed onto the FPGA and results streamed back, with the host undertaking the in-direct memory accesses and result accumulations.}
\label{fig:streaming_design}
\end{figure}

On the other hand, whilst the FPGA is actively computing the host CPU is idle. Therefore a question was whether it would be beneficial to exploit this fact by moving these irregular memory access and result accumulations onto the CPU. This would enable the CPU to stream required data to the matrix assembly engine for each element and for the engine's results to be streamed back and processed, keeping the FPGA design continually busy with the CPU concurrently undertaking those aspects less suited to the FPGA. Whilst this might sound like an obvious use of a streaming interface, such as AXIS, between the host and kernel, the Alveo U280 shell only supports DMA \cite{u280}. Therefore in order to achieve this streaming approach we needed to implement it using AXI4 data transfers via the external HBM2. Figure \ref{fig:streaming_design} illustrates an architectural view of this approach, where we have an additional HLS-based IP block which read data from the on-board but off-chip HBM2 memory and then streams this via AXIS to the matrix assembly engine. There are two output HLS-based IP blocks which receive results streamed from the engine and store these in the HBM2. 

Our approach operates with dynamically sized chunks, where the host CPU executes the \emph{calculate\_transients} procedure from Listing \ref{lst:input_output}, extracting data from \emph{veloc} and \emph{coord} input arrays for each of the chunk's elements into \emph{elvel} and \emph{elcod} respectively. When a chunk of data, containing \emph{elvel} and \emph{elcod}, is ready it is then transferred to the FPGA via PCIe and the \emph{Stream elvel and elcod in} kernel started. This kernel retrieves the \emph{elvel} and \emph{elcod} values for each element from HBM2 and streams these to the matrix assembly engine. Handling results data is the opposite, where the \emph{stream elrbu out} and \emph{stream eldtrho and elmurho out} kernels are scheduled for each chunk, and as this data arrives it is stored in HBM2 and then transferred from the FPGA to the host via PCIe and processed. We marshal these activities via OpenCL events, where not only is the data transfer from the FPGA to the host scheduled a priori and then dependent on the streaming output IP blocks completing for a specific chunk, but furthermore the OpenCL \emph{setCallback} function is used to schedule the execution of a result unpacking function. This is effectively the \emph{perform\_assembly\_in\_global\_system} function and runs on the CPU when data migration of results to the host has completed, accumulating the \emph{elrbu}, \emph{eldtrho}, and \emph{elmurho} values into \emph{rhsid}, \emph{dt\_rho\_nsi}, and \emph{mass\_rho\_nsi} arrays respectively.

The idea is that, as the entire data-set has been split into chunks, then the host will be busy loading the next chunk, whilst the FPGA is running with the current chunk, and the host also processing result data from the previous chunk. Furthermore, to maximise performance, following guidance in \cite{apchain}, our input and output streaming  HLS-based IP blocks use the \emph{ap\_ctrl\_chain} interface pragma to apply back-pressure and enable multiple kernel executions to be overlapped and run in a pipelined fashion.

The performance of this approach is illustrated in Table \ref{fig:kernel_performance_streaming}, which details the total execution time only, as data transfers to and from the FPGA via PCIe and kernel execution are overlapped so it no longer makes sense to include these two individual metrics. The entry \emph{Initial streaming approach} illustrates the performance benefits, resulting in around a three times increase in performance on the FPGA which was now achieving around 63\% the CPU performance. However, at this point, we realised that the matrix assembly engine was still being starved of work as it needed 12 double precision numbers from \emph{elvel} and \emph{elcod} per element, but the streaming IP block was only reading data from HBM2 memory in widths of 512 bit effectively meaning that there were only 8 values available per cycle. To enable the streaming of a complete \emph{elvel} and \emph{elcod} set of values per cycle, we reorganised the data such that each variable was split across two HBM2 memory banks\footnote{On the Alveo U280 HBM2 is split into thirty two memory banks each of size 256MB. There are sixteen memory controllers, each with a dedicated channel to both banks it controls.} and AXI4 kernel ports. Hence there are four AXI4 connections between the \emph{Stream elvel and elcod in} IP block of Figure \ref{fig:streaming_design} and the HBM2, two for each input variable. An element's input of each variable is therefore split across these two AXI4 ports and bank of HBM2, with the lower 6 double precision numbers in the first part and higher 6 numbers in the second part. To keep the code simple we pad the rest of the 512 bit chunk, so there is 64 bit of data padded per bank per element, whilst this does result in some wasted memory it makes the device code significantly simpler. 


This optimisation enabled the Alya engine to be fed with data each cycle, and the performance is reported in Table \ref{fig:kernel_performance_streaming} by the \emph{Data streamed each cycle} entry. It can be seen that this double the performance on the FPGA, and for the first time we were out-performing the 24-core Xeon Platinum CPU. 

As described above, we were using the OpenCL \emph{setCallback} API call to execute our results handling function which accumulates results from the FPGA to their appropriate locations in the data arrays. However these functions were executing sequentially on a single CPU core, potentially resulting in a bottleneck. Therefore we parallelised these using threading, with a thread issued for each result accumulation function execution. This meant that result accumulations could occur for multiple data chunks concurrently, and whilst it is fairly simplistic, for instance a new thread is created each time and there is no guard against oversubscribing threads to cores for large number of chunks, it almost doubled the performance of matrix assembly on the FPGA. This is reported by the entry \emph{Threaded result handling} in Table \ref{fig:kernel_performance_streaming}, and now we were outperforming the 24-core CPU by over two times.

\begin{table}[h]

 \centering
\begin{tabular}{ | c c c | }
\hline
\textbf{Description} & \makecell{\textbf{Execution} \\ \textbf{time (ms)}} & \textbf{\% CPU performance}\\\hline
24 cores Xeon CPU & 61.72 & - \\
Previous non-streaming FPGA & 284.04 & 21.73\%\\
Initial streaming approach & 97.91 & 63.04\%\\
Data streamed each cycle & 48.19 & 128.08\%\\
Threaded result handling & 26.67 & 231.42\%\\
\hline
\end{tabular}
\caption{FPGA matrix assembly performance for Sphere 100K benchmark on Alveo U280 for our data streaming approach. Compares against the previous non-data streaming FPGA design and 24-core Xeon Platinum CPU}
\label{fig:kernel_performance_streaming}
\end{table}

By moving those aspects less suited to the FPGA, namely the indirect memory accesses and results accumulations, onto the CPU cores we obtained a speed up of around 11 times on the FPGA compared with our previous non-streaming FPGA design. Therefore clearly it was beneficial to take advantage of the idle CPU cores and, whilst the Alveo U280 did not explicitly support a host-FPGA AXIS streaming interface, it was possible to develop an approach which successfully followed the general idea. Moreover, the optimisations described in this section have resulted in an FPGA matrix assembly engine which is over 600 times faster than the initial dataflow design detailed in Table \ref{fig:kernel_performance}.

\subsection{Scaling to multiple FPGA engines}

Performance described thus far has focused on running a single Alya incompressible flow matrix assembly engine, and it is possible to scale this up to multiple engines. A challenge was that the current engine was using 97\% of an SLR's DSP slices (there are three SLRs on the U280). With this utilisation it was only possible to fit two engines on the FPGA, as any more would result in routing errors. However, it was observed that the \emph{calculate\_cartesian\_derivatives} function, which only accounts for around 5\% of the overall matrix assembly runtime required 33\% of the DSP slices. Therefore we moved this function off the FPGA onto the CPU, with results from this calculation streamed in to the rest of the engine. This follows the same idea adopted for streaming data described in Section \ref{sec:improve_read_write}, where we are exploiting the otherwise under utilised CPU to share the workload and make better use of the FPGA. 

We adopted a similar streaming approach to that of Figure \ref{fig:streaming_design}, still with a single input and two output HLS-based streaming IP blocks, but these serviced multiple matrix assembly engines. Each engine runs concurrently and processes different elements. The streaming IP blocks where modified slightly to maximise performance, where there are two AXI4 links into the input block per variable, for each engine. The idea was that these can all be read concurrently and therefore all engines continually fed with data. The output blocks read from all engines concurrently per cycle and pack the data into 512 bit wide writes, but still with a single AXI4 channel per variable. Initially, running calculations of cartesian derivatives on the host increased the runtime by around 30\%, and we then threaded the processing of each chunk to run concurrently. The threading is different to that of threading the handling of results, because there is no benefit in running the handling of each input chunk in parallel as it does not reduce the time for the first to be ready, instead with them simply queuing up.

Instead we applied threading via OpenMP inside the processing of each chunk, performing the cartesian derivative calculations for a specific chunk in parallel across the CPU cores. This sped up the calculations involved for each chunk, meaning they were ready sooner and improved performance, resulting in a negligible overhead compared with undertaking the \emph{calculate\_cartesian\_derivatives} function on the FPGA.

\section{Performance and power comparison}
\label{sec:performance}
\begin{figure}[h]
\centering
\includegraphics[scale=0.43]{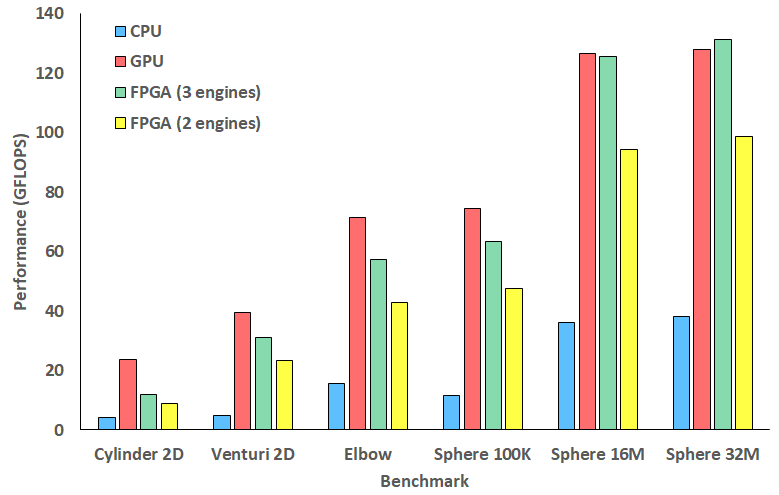}
\caption{Performance comparison between CPU, GPU, three kernels on the FPGA (with cartesian derivatives calculation on CPU) and two kernels on the FPGA (all calculations on the FPGA). Higher is better.}
\label{fig:performance}
\end{figure}

We undertook performance and power consumption experiments of our approach against the 24-core Xeon Platinum (Cascade Lake) 8260M CPU (threaded via OpenMP) and Nvidia V100 GPU (using the existing Alya OpenACC implementation \cite{oyarzun2021performance}). We explored two FPGA configurations, the two engine approach where the \emph{calculate\_cartesian\_derivatives} function is resident on the FPGA, and the three engine approach where that function is running on the host CPU instead. All FPGA experiments are running at a clock frequency of 300MHz. In this section we report performance in GFLOPS across the benchmarks described in Table \ref{tbl:benchmark_details}, and this is illustrated by Figure \ref{fig:performance}. It can be seen that CPU performance is consistently the lowest performing of three technologies, delivering especially poor performance for the smaller benchmarks. Whilst the CPU's performance improves for two larger benchmarks, it still falls considerably short of both the GPU and FPGA. The GPU outperforms the FPGA until the larger two benchmark sizes, where the three engine FPGA design performs comparably and even slightly out performs the GPU for the Sphere 32M benchmark. As would be expected, the three engine FPGA approach outperforms the two engine approach consistently.

\begin{figure}[h]
\centering
\includegraphics[scale=0.43]{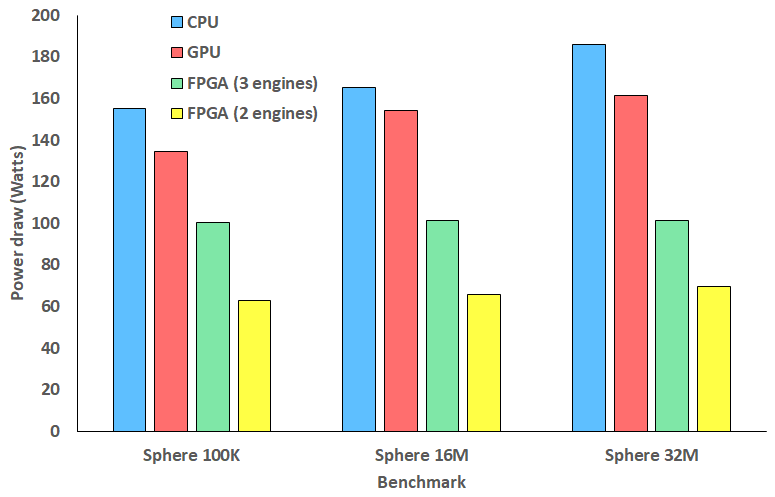}
\caption{Power draw comparison between CPU, GPU, three kernels on the FPGA (with cartesian derivatives calculation on CPU) and two kernels on the FPGA (all calculations on the FPGA). Lower is better.}
\label{fig:power}
\end{figure}

Figure \ref{fig:power} illustrates power usage in Watts between different hardware technologies running our matrix assembly kernel. Power was captured on the CPU using RAPL, NVIDIA-SMI for the GPU, and XRT for the FPGA, with the average power draw reported in Figure \ref{fig:power}. Due to the very short runtimes of smaller benchmarks, we only report power for larger benchmarks where the runtime was sufficient to gather a reliable measurement. For the FPGA the combined FPGA and CPU power draw is reported because activities are shared. 

It can be seen from Figure \ref{fig:power} that the CPU draws the most power, with the GPU drawing slightly less but still greater than 160 Watts for the largest benchmark. The FPGA power draw is significantly less than the other technologies, with three engines drawing 100 Watts for the largest benchmark and two engines 70 Watts. The majority of this difference in FPGA power draw comes from the CPU, which draws more power for the three engine approach as the cartesian derivative calculations are on the CPU. By contrast, even though the two engine approach still processes input, and result data accesses and accumulations on the CPU, this is significantly less workload than the threading floating point operations required by the cartesian derivatives calculations, and-so power draw is much less. It is interesting that, for the three engine approach, the power does not increase for larger benchmarks. As the CPU will be undertaking more computation one would assume it would draw more power as it is below the TDP. The fact that it does not indicates that this workload on the CPU is trivial, which corresponds with details in Table \ref{tbl:matrix_assembly_details}, and likely stop and start.

\begin{figure}[h]
\centering
\includegraphics[scale=0.43]{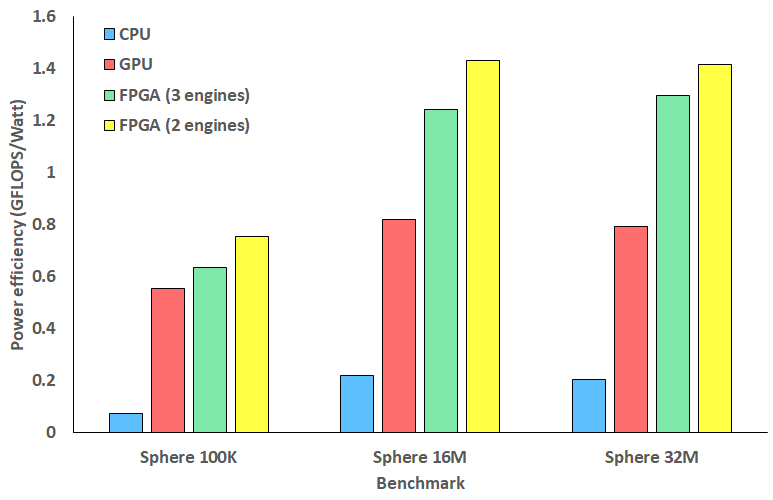}
\caption{Power efficiency comparison between CPU, GPU, three kernels on the FPGA (with cartesian derivatives calculation on CPU) and two kernels on the FPGA (all calculations on the FPGA). Higher is better.}
\label{fig:power_efficiency}
\end{figure}

Figure \ref{fig:power_efficiency} illustrates the power efficiency, in GFLOPS/Watt, for each of the three largest benchmarks. It can be seen that the CPU is significantly worse than the other two technologies with the GPU falling short of the FPGA. Interestingly both FPGA configurations are fairly comparable, with the increased performance of the three engine design offsetting the reduced power draw of the two engine approach.

\section{Conclusions and further work}
\label{sec:conclusions}
In this paper we have explored the porting of Alya's incompressible flow matrix assembly kernel onto a Xilinx Alveo U280 FPGA. Starting with a naive dataflow design, it required multiple optimisation steps to fully suit the FPGA, with our final single engine achieving over double the performance of the CPU. This was delivered by ensuring that data could continually flow through the dataflow engine, keeping all parts running concurrently and included a mixture of improvements to the structure of the algorithms and also mapping the most appropriate parts of the workload between the CPU and FPGA. 

We went further to explore mixing calculations on the CPU and FPGA to enable us to reduce the per-engine DSP usage, and enable an increase in the number of engines on the FPGA. This is an interesting example of the FPGA and CPU working to compliment each other, and is likely to become more important as Xilinx's next generation Versal cards are released which combine the reconfigurable fabric with a multi-core CPU and simple vector-style AI engines on-chip.

We compared the performance, power draw, and power efficiency against a 24-core Xeon Platinum (Cascade Lake) 8260M CPU and Nvidia V100 GPU. Our FPGA approach considerably out-performed the CPU and our three-engine version performed comparably against the GPU, especially for larger benchmarks. The benefit of the FPGA is stark when one considers power, where the FPGA configurations drew significantly less power than the CPU and GPU and hence delivered greater power efficiency especially for the Sphere 16M and 32M benchmarks. Due to the high power consumption of the CPU, our performance comparison also illustrated a trade-off when deciding whether to mix floating point calculations on the CPU and FPGA. Such mixing enabled scaling to three engines, which improved performance, however it also drew a third more power compared to the two engine approach which does not mix. In this case however it was observed that when it came to power efficiency the performance and power draw largely cancelled each other out, with the FPGA configurations delivering very similar power efficiency characteristics.

For further work we believe that next-generation future Versal architecture will be interesting to target, especially the AI engines which accelerate single precision floating point and fixed point arithmetic. This will likely substantially reduce the DSP usage and these AI engines running at 1GHz will likely further increase performance. Additionally there is more work to be done in exploring the mixing of workloads across the CPU and FPGA, potentially using CPU soft-cores such as the MicroBlaze in the fabric to run some of these constituent workloads which are especially suited to the CPU only.

We conclude that the use of FPGAs is beneficial for engineering HPS simulations, Alya in particular, and is an important future hardware technology that should be considered as part of exascale supercomputers. The fact that our design was able to out-perform the CPU and perform comparably against the GPU demonstrates that FPGAs play an important part in ameliorating non-compute overheads, where GPUs clearly have much greater raw floating-point capability but can be limited by overheads imposed by the general purpose architecture. Furthermore our FPGA approach drew substantially less power than the other technologies, resulting in much greater power efficiency. However there is an important note of caution, even though we had adopted what seemed like a sensible initial dataflow design, this still required in-depth optimisation and exploration to make most effective use of the technology, and we believe that these dataflow algorithmic investigations are an important activity to be undertaken by the community to fully mature FPGAs for next-generation HPC workloads.

\section*{Acknowledgement}

The authors would like to thank the ExCALIBUR H\&ES FPGA testbed and Xilinx XACC program for access to compute resource used in this work. This work was funded under the EU EXCELLERAT CoE, grant agreement number 823691.

\bibliographystyle{IEEEtran}
\bibliography{references}

\appendices

\section{Artifact Description Appendix}
\label{sec:ad}

\subsection{Description}

\subsubsection{Check-list (artifact meta information)}

{\small
\begin{itemize}  
  \item {\bf Program: } C++ and Fortran
  \item {\bf Compilation: }GCC version 8.3 with -O3, Vitis version 2021.1. On the GPU we used the Nvidia Compiler version  20.9.
  \item {\bf Data set: }Runs were based on the \emph{Cylinder 2D}, \emph{Venturi 2D}, \emph{Elbow}, and \emph{Sphere 16M} Alya benchmarks. As described in the paper, we added an additional two benchmarks (\emph{Sphere 100K} and \emph{Sphere 32M}) based on the Sphere benchmark to add some additional element size variety.
  \item {\bf Run-time environment: } A variety of machines were used for comparison, all running Linux. For the Alveo U280 the latest environment at the time of writing was used (XRT 202110.2.11.634, xdma and xdma-dev 201920.3 deployment and development target platforms)
  \item {\bf Hardware: } We used a Xilinx Alveo U280 for the FPGA runs, this is hosted by a system with a Xeon Platinum Skylake (8170) and 192GB RAM. For CPU comparison runs we ran on a Xeon Platinum Cascade Lake (8260M) processor with 192GB RAM. For GPU runs we ran on the Cirrus tier-2 UK HPC machine, which provides a NVIDIA Tesla V100-SXM2-16GB (Volta) GPU, hosted by two Intel Xeon Gold Cascade Lake (6248) CPUs and 384 GB RAM.
  \item {\bf Binary: } Alya CPU versions require MPI. Xilinx Vitis library is required to synthesise the kernel and generate the bitstream.
  \item {\bf Execution: } We built and executed all executables on Linux.
  \item {\bf Output: } Alya produces detailed logs that provide performance measurements and we also added in additional timing and manually calculated to ensure that the reported value was correct.
  \item {\bf Publicly available?: }No, Alya is commercial software available under licence
\end{itemize}
}

\subsubsection{Hardware dependencies}
Any machine running Linux with appropriate Alveo U280 FPGA PCIe card installed
\subsubsection{Software dependencies}
The latest version of Alya, GCC version 8.3, the support libraries installed for the board and the Vitis platform. 
\subsubsection{Datasets}
Runs were based on the Alya benchmark suite which is also provided as part of the Alya distribution
\subsection{Installation}
We synthesised our kernel using Vitis HLS via the \emph{v++} command. It is also possible to synthesise directly via the HLS IDE, and this was useful for steps which involved leveraging the analysis pane. The \emph{v++} command was then used to link, which assembles the shell and calls out to Vivido to generate the bitstream. The host code was written in OpenCL (the appropriate libraries ship with Vitis) and launching our bitstream simply involved executing the host code, which via the appropriate OpenCL calls programmed the device as appropriate. 

\subsection{Experiment workflow}
\begin{enumerate}
\item Develop the appropriate HLS kernel 
\item Use Vitis HLS to synthesise this and generating corresponding \emph{.xo} files
\item Use Vitis HLS in linking mode to generate the bitstream \emph{.xclbin} file
\item Compile the host OpenCL code using GCC
\item Execute the host code, which will launch the bitstream
\item Optionally, enable profiling and after the run use Vitis Analyser to explore this information.
\end{enumerate}

\subsection{Evaluation and expected result}
We compared our results against the existing CPU and GPU version of Alya. On the CPU this ran across all 24 cores of the 8260M and was threaded via OpenMP. On the GPU Alya provides an OpenACC implementation of the matrix assembly kernel which is the code that was used. All calculated values have been checked at the element level to ensure that they are producing consistent results, and all performance and power results reported in this paper are averaged over three runs.

\subsection{Experiment customization}
It is, of course, possible to experiment with the benchmarks and use these to run different system sizes, for instance modifying the number of elements or points per kernel. On an Alveo U280 the maximum number of kernels we could fit was three, and with a larger FPGA such as the future Versal architecture, then this could be scaled up further potentially.

\section{Artifact Evaluation}

\subsection{Results Analysis Discussion}
In the host code we use OpenCL's profiling capability which provides microsecond resolution timings for event (kernel execution) starting and ending. We also implemented manual timing via the \emph{gettimeofday} call, to provide a second timing comparison point and ensure what OpenCL reported was correct (both approaches to timing matched very closely.) All calculated values were checked for consistency between the FPGA and CPU versions to ensure that they are calculating the same quantities and we were undertaking a fair experiment. For all experiments runtimes were averaged over at-least three runs, and power consumption figures were reported by XRT for the FPGA, RAPL for the CPU, and \emph{nvidia-smi} on the GPU.

\end{document}